\documentclass[11pt]{article}
\usepackage{amsmath,amsfonts,amssymb,cite}
\newcommand{\be}{\begin{equation}}
\newcommand{\ee}{\end{equation}}
\newcommand{\ba}{\begin{eqnarray}}
\newcommand{\ea}{\end{eqnarray}}

\newcommand{\la}[1]{\label{#1}}

\def\gl#1{(\ref{#1})}

\date{}
\begin{document}
\title{Minimal Realizations of Supersymmetry\\ for Matrix Hamiltonians}
\author{Alexander A. Andrianov,\,
Andrey V. Sokolov\\ \\{\it Department of
Theoretical Physics, Saint-Petersburg State University,}\\{\it Ulianovskaya ul., 1, Petrodvorets,
Saint-Petersburg 198504, Russia}}
\maketitle
\abstract{The notions of weak and strong minimizability of a matrix intertwining operator are introduced.
Criterion of strong minimizability of a matrix intertwining operator is revealed.
Criterion and sufficient condition of existence of a constant symmetry matrix for a matrix Hamiltonian are presented.
A method of constructing of a matrix Hamiltonian with a given constant symmetry matrix in terms of a set of arbitrary scalar functions and eigen- and associated vectors of this matrix is offered.
Examples of constructing of $2\times2$ matrix Hamiltonians with given symmetry matrices for the cases of different structure of Jordan form of these matrices are elucidated.}

\section{Introduction}
The matrix models with supersymmetry appear in Quantum Mechanics in several areas: in particular, for spectral design of potentials describing multichannel scattering and the motion of spin particles in external fields. The different cases of such models are considered in \cite{anio88,acd88,acd90-1,anio91,aisv91,canio92,canio93,takahiro93,hgb95,basp97,lrsfv98,dc99,tr99,iknn06,fmn10,coja14} and their systematic study was undertaken in \cite{acni97,gove98,sakhnovich,ione03,sampe03,suzko05,pepusa11,tanaka11,sokolov13,pupasov13,nk11,nika11,ni12,ka12,ka13,cinn04,matrix14} (see also the recent reviews \cite{pup14,andrianovioffe}). In \cite{acni97}  intertwining of matrix Hermitian Hamiltonians by $n\times n$ first-order and $2\times 2$ second-order matrix differential operators was investigated and the corresponding supersymmetric algebras were constructed. The main result of \cite{gove98} is the formulae that help to build  a modified   $n\times n$ matrix Hamiltonian for a given $n\times n$ matrix (non-Hermitian, in general) Hamiltonian. The intertwining operator for this purpose is given by a $n\times n$ matrix linear differential operator of arbitrary order with the identity matrix coefficient at derivative $d/dx$ in the highest degree that intertwines these Hamiltonians. The systematic study of intertwining relations for $n\times n$ matrix non-Hermitian, in general, one-di\-mensional Hamiltonians has been performed in \cite{sokolov13,matrix14} with intertwining realized by $n\!\times\! n$ matrix linear differential operators with nondegenerate coefficients at $d/dx$ in the highest degree. Some methods of constructing of $n\!\times\! n$ matrix intertwining operator of the first order in derivative and of general form were proposed and their interrelations were examined.

In the one-dimensional QM with scalar Hamiltonians the isospectral transformations generally lead to a Nonlinear (Polynomial \cite{ais,acdi95}) SUSY algebra. In respect to the  SUSY partners there might be an infinite number of intertwining operators which provide the same pair of potentials but a different SUSY algebras \cite{anso03,ancaso07}. These intertwining operators differ in factors which are functions of the Hamiltonians themselves. Thus for intertwining operators and the SUSY algebra itself the problem  arises  to minimize the order of their differential representation --- minimizability problem. For some classes of potentials  minimized algebras may exist which contain the same partner Hamiltonians but different sets of intertwining operators. In this case the hidden symmetry operators appear  \cite{anso03} being built of products of intertwining operators from different algebras. In the case of scalar Hamiltonians the number of independent minimized algebras cannot exceed two. The similar program has not been realized for matrix SUSY partner Hamiltonians although the important steps in this direction were done in \cite{sokolov13}.

In the present paper the following new results are elaborated:
\begin{itemize}
\item The notions of weak and strong minimizability of a matrix intertwining operator are introduced.
\item Criterion of strong minimizability of a matrix intertwining operator is revealed.
\item Criterion and sufficient condition of existence of a constant symmetry matrix for a matrix Hamiltonian are presented.
\item A method of constructing of a matrix Hamiltonian with a given constant symmetry matrix in terms of a set of arbitrary scalar functions and eigen- and associated vectors of this matrix is offered.
\item Examples of constructing of $2\times2$ matrix Hamiltonians with given symmetry matrices for the cases of different structure of Jordan form of these matrices are elucidated.
\end{itemize}
 The basic notations for intertwining operator algebra are defined in Sec.2.
 In Sec.3 we explain the motivation for introduction of the notions of weak and strong minimizabilities and give their definitions. The previous results on minimizability in scalar case are briefly formulated: in particular, the criterion of complete minimizability for a scalar intertwining operator from \cite{anso03,ancaso07}. At the end of this section the criterion of complete weak minimizability is given for a matrix intertwining operator from \cite{sokolov12,sokolov13} and the criterion of partial strong minimizability from the right for a matrix intertwining operator is presented. In Sec.4 the sufficient condition of existence of a constant symmetry matrix for a matrix Hamiltonian is found. This condition provide us with the opportunity to receive the useful in practice formula  for constructing of a matrix Hamiltonian with a given constant symmetry matrix. The two examples of constructing of arbitrary $2\times2$ matrix Hamiltonian with a given symmetry matrix $A$ are presented. In the first of these examples there are two different eigenvalues for the matrix $A$ and in the second of these examples normal (Jordan) form of $A$ is a Jordan block. In Conclusion we outline the perspectives for further studies of the criteria of minimizabilities and hidden symmetries induced by extended SUSY algebras.

\section{Basic definitions and notation\la{sec2}}

Let us consider two  $n\times n$ matrix Hamiltonians defined on the entire axis,
\[H_+=-I_n\partial^2+V_+(x),\quad H_-=-I_n\partial^2+V_-(x),\qquad\partial\equiv{d/{dx}},\la{h+h-2.1}\]
where $I_n$ is the identity matrix and $V_+(x)$ and $V_-(x)$ are square matrices, all elements of which are sufficiently smooth and, in general, complex-valued functions. We suppose that these Hamiltonians are {\it intertwined} by a matrix linear differential operator $Q_N^-$, so that
\begin{equation} Q_N^-H_+=H_-Q^-_N,\qquad Q_N^-=\sum\limits_{j=0}^NX^-_j(x)\partial^j,\la{splet}\end{equation}
where $X^-_j(x)$, $j=0$, \dots, $N$ are also square matrices of $n$-th order, all elements of which are sufficiently smooth and, in
general, complex-valued functions.

It is not hard to check that intertwining \gl{splet} leads to the following consequences:
\begin{equation}X_N^-(x)={\rm Const},\qquad V_-(x)=X^-_NV_+(x)(X^-_N)^{-1}+2X^{-\,\prime}_{N-1}(x)(X^-_N)^{-1},\la{vmp2.1}\end{equation}
where and in what follows we restrict ourselves by the case $\det X_N^-\ne0$.

Let's elucidate the structure of intertwining operator kernel and define the  transformation vector-functions.
By virtue of the intertwining the kernel of $Q_N^-$ is an invariant subspace for~$H_+$:
\[H_+\ker Q_N^-\subset\ker Q_N^-.\]
Hence, for any basis $\Phi^-_1(x)$, \dots, $\Phi^-_d(x)$ in $\ker Q_N^-$, $d=\dim\ker Q_N^-=nN$ there is a constant square matrix $T^+\equiv\|T_{ij}^+\|$ of $d$-th order such that
\[H_+\Phi^-_i=\sum\limits_{j=1}^dT^+_{ij}\Phi^-_j,\qquad i=1,\ldots,d.\la{tm}\]

A basis in the kernel of an intertwining operator $Q_N^-$ in which the matrix $T^+$ has a normal (Jordan) form is called a {\it
canonical basis}. Elements of a canonical basis are called a {\it transformation vector-functions}.

If a Jordan form of the matrix $T^+$ has block(s) of order higher than one, then the corresponding canonical basis contains not only formal vector-eigenfunctions of $H_+$ but also formal associated vector-function(s) of $H_+$ which are defined as follows \cite{naim}.

A finite or infinite set of vector-functions $\Phi_{m,i}(x)$, $i=0$, 1, 2, \dots\, is called a {\it chain of formal associated vector-functions} of $H_+$ for a spectral value $\lambda_m$ if
\[H_+\Phi_{m,0}=\lambda_m\Phi_{m,0},\quad\Phi_{m,0}(x)\not\equiv0,
\qquad (H_+-\lambda_mI_n)\Phi_{m,i}=\Phi_{m,i-1},\quad i=1,2,3,\ldots\,.\]

\section{Weak and strong minimizability of a matrix intertwining operator\la{sec3}}

\vskip-0.2pc

Let us introduce the notions of weak and strong minimizabilities and give their definitions.

\vskip0.5pc

It is evident that if to multiply $Q_N^-$ by a polynomial of the Hamiltonian,
\[Q_N^-\Big[\sum\limits_{l=0}^LA_lH_+^l\Big]\qquad \Big(\Big[\sum\limits_{l=0}^{L}A_lH_-^l\Big]Q_N^-\Big),\]  where $A_l$, $l=0$, \dots, $L$ are either a complex numbers or constant symmetry matrices for $H_+$ ($H_-$), then such products are again an intertwining operators for the same Hamiltonians:
\[\Big\{Q_N^-\Big[\sum\limits_{l=0}^LA_lH_+^l\Big]\Big\}H_+=Q_N^-H_+\Big[\sum\limits_{l=0}^LA_lH_+^l\Big]=H_-\Big\{Q_N^-\Big[\sum\limits_{l=0}^LA_lH_+^l\Big]\Big\}\]
\[\left(\Big\{\Big[\sum\limits_{l=0}^LA_lH_-^l\Big]Q_N^-\Big\}H_+=\Big[\sum\limits_{l=0}^LA_lH_-^l\Big]H_-Q_N^-=H_-\Big\{\Big[\sum\limits_{l=0}^LA_l H_-^l\Big]Q_N^-\Big\}\right).\]
Thus, the question arises about possibility to simplify an intertwining operator by separation from it a superfluous factor which is a polynomial of the corresponding Hamiltonian.

Let us now present the formal definitions of weak and strong minimizability of a matrix intertwining operator.

\vskip0.5pc

\noindent{\bf Definition 1.} An intertwining operator $Q_N^-$ is called {\it weakly minimizable} if this operator can be represented in the form
\[Q_N^-=P_M^-\Big[\sum\limits_{l=0}^La_lH_+^l\Big]\equiv \Big[\sum\limits_{l=0}^La_lH_-^l\Big]P_M^-,\]
\begin{equation}a_l\in\Bbb C,\qquad l=0, \ldots,L,\qquad a_L\ne0,\qquad1\leqslant L\leqslant {N\over2},\la{wmin}\end{equation}
where $P_M^-$ is an $n\times n$ matrix linear differential operator of $M$-th order, $M=N-2L$ that intertwines the Hamiltonians $H_+$ and $H_-$, so that $P_M^-H_+=H_-P_M^-$. Otherwise, the operator $Q_N^-$ is called {\it weakly non-minimizable}.

\vskip0.5pc

\noindent{\bf Definition 2.} An intertwining operator $Q_N^-$ is called {\it strongly minimizable from the right $($from the left$)$} if this operator can be represented in the form
\begin{equation}Q_N^-=P_M^-\Big[\sum\limits_{l=0}^LA_lH_+^l\Big]\qquad \Big(Q_N^-=\Big[\sum\limits_{l=0}^LA_lH_-^l\Big]P_M^-\Big),\la{smin}\end{equation}
where $A_l$, $l=0$, \dots, $L$, $1\leqslant L\leqslant N/2$ are a symmetry $n\times n$ matrices for $H_+$ ($H_-$), $\det A_L\ne0$ and $P_M^-$ is an $n\times n$ matrix linear differential operator of $M$-th order, $M=N-2L$ that intertwines the Hamiltonians $H_+$ and $H_-$, so that $P_M^-H_+=H_-P_M^-$. Otherwise, the operator $Q_N^-$ is called {\it strongly non-minimizable from the right $($left$)$}.

\vskip0.8pc

\noindent{\bf Remark 1.} In the scalar case $n=1$ weak and strong minimizabilities are equivalent obviously. Thus. we shall speak below in the scalar case $n=1$ about minimizability without additional specification ``weak'' or ``strong''.

For comparison with our new results on strong minimizability of a matrix intertwining operator  we present the previously formulated criterion of complete minimizability for a scalar intertwining operator \cite{anso03,ancaso07} and the criterion of complete weak minimizability for a matrix intertwining operator \cite{sokolov12,sokolov13}. The word ``complete'' means here that we consider such a minimization of an intertwining operator $Q_N^-$ that the corresponding residual intertwining operator $P_M^-$ in \gl{wmin} is (weakly) non-minimizable.\\

\noindent{\bf Criterion of complete minimizability of a scalar intertwining operator.}

\noindent {\it In the scalar case $n=1$ an intertwining ope\-ra\-tor~$Q_N^-$ can be represented in the form
\[Q_N^-=P_M^-\prod\limits_{l=1}^s(\lambda_l-H_+)^{ k_l},\qquad\lambda_l\in{\Bbb C},\,\, k_l\in\Bbb N,\,\,
l=1, \ldots, s,\quad\lambda_l\ne \lambda_{l'}\Leftrightarrow l\ne l',\la{minim8}\]
where $P_M^-$ is a non-minimizable linear differential operator of the $M$-th order that intertwines the Hamiltonians $H_+$ and $H_-$, so that $P_M^-H_+=H_-P_M^-$,

\vskip0.5pc

\noindent if and only if

\renewcommand{\labelenumi}{\rm{(\theenumi)}}
\begin{enumerate}
\item all numbers $\lambda_l$, $l=1$, \dots, $s$ belong to the spectrum of the matrix $T^+$
and there are no equal numbers between them$;$
\item there are $2$ Jordan blocks in a normal $($Jordan$)$ form of the matrix $T^+$ for any eigenvalue from the set $\lambda_l$, $l=1$, \dots, $s;$
\item there are no $2$ Jordan blocks in a normal $($Jordan$)$ form of $T^+$ for any eigenvalue of this matrix that does not belong to the set $\lambda_l$, $l=1$, \dots, $s;$
\item $ k_l$ is the minimal of the orders of Jordan blocks corresponding to the eigenvalue $\lambda_l$ in a normal $($Jordan$)$ form of the matrix $T^+$, $l=1$, \dots, $s.$
\end{enumerate}}

\vskip1pc

\noindent{\bf Criterion of complete weak minimizability of a matrix intertwining operator.}

\noindent{\it A~matrix intertwining operator $Q_N^-$ can be represented in the form
\[Q_N^-=P_M^-\prod\limits_{l=1}^s(\lambda_lI_n-H_+)^{ k_l},\qquad\lambda_l\in{\Bbb C},\,\, k_l\in\Bbb N,\,\,
l=1, \ldots, s,\quad\lambda_l\ne \lambda_{l'}\Leftrightarrow l\ne l',\la{minim8}\]
where $P_M^-$ is a non-minimizable matrix linear differential operator of the $M$-th order that intertwines the Hamiltonians $H_+$ and $H_-$, so that $P_M^-H_+=H_-P_M^-$,

\vskip0.5pc

\noindent if and only if

\renewcommand{\labelenumi}{\rm{(\theenumi)}}
\begin{enumerate}
\item all numbers $\lambda_l$, $l=1$, \dots, $s$ belong to the spectrum of the matrix $T^+$
and there are no equal numbers between them$;$
\item there are $2n$ Jordan blocks in a normal $($Jordan$)$ form of the matrix $T^+$ for any eigenvalue from the set $\lambda_l$, $l=1$, \dots, $s;$
\item there are no $2n$ Jordan blocks in a normal $($Jordan$)$ form of $T^+$ for any eigenvalue of this matrix that does not belong to the set $\lambda_l$, $l=1$, \dots, $s;$
\item $ k_l$ is the minimal of the orders of Jordan blocks corresponding to the eigenvalue $\lambda_l$ in a normal $($Jordan$)$ form of the matrix $T^+$, $l=1$, \dots, $s.$
\end{enumerate}}

 Now let us give the criterion of partial strong minimizability from the right for a matrix intertwining operator. The word ``partial'' means here that we consider such minimization of a matrix intertwining operator $Q_N^-$ that the corresponding residual intertwining operator $P_M^-$ in \gl{smin} is not necessarily strongly non-minimizable from the right. As well we present in this section the interesting corollary of the mentioned above criterion that contains the criterion for a constant $n\times n$ matrix to be a symmetry matrix for an $n\times n$ matrix Hamiltonian.\\

\noindent{\bf Criterion of partial strong minimizability from the right of a matrix intertwining operator.}

\noindent{\it A~matrix intertwining operator $Q_N^-$ can be represented in the form
\[Q_N^-=P_{N-2}^-(A-H_+),\la{minim8c}\]
where $A$ is a constant $n\times n$ matrix and $P_{N-2}^-$ is a matrix linear differential operator of the order $N-2$ that intertwines $H_+$ and $H_-$, so that $P_M^-H_+=H_-P_M^-$, and the matrix $A$ is a symmetry matrix for the Hamiltonian $H_+$, $[H_+,A]=0$,

\vskip0.5pc

\noindent if and only if

\vskip0.5pc

\noindent the kernel $\ker Q_N^-$ contains double set of associated vector-functions
\[\Phi_{ial}(x),\quad\Psi_{ial}(x),\qquad i=1,\ldots, m,\quad a=1,\ldots,g_i,\quad l=0,\ldots, \nu_{ia}-1,\hfill\] \[\sum\limits_{i=1}^m\sum\limits_{a=1}^{g_i}\nu_{ia}=n,\qquad \nu_{i1}\geqslant\nu_{i2}\geqslant\ldots\geqslant\nu_{i,g_i},\forall i\]

\noindent such that

\renewcommand{\labelenumi}{\rm{(\theenumi)}}
\begin{enumerate}
\item
\[H_+\Phi_{ia0}=\lambda_i\Phi_{ia0},\qquad(H_+-\lambda_iI_n)\Phi_{ial}=\Phi_{ia,l-1},\]
\[H_+\Psi_{ia0}=\lambda_i\Psi_{ia0},\qquad(H_+-\lambda_iI_n)\Psi_{ial}=\Psi_{ia,l-1},\]
\[\lambda_i\in{\Bbb C},\qquad i=1,\ldots, m,\quad a=1,\ldots,g_i,\quad l=1,\ldots, \nu_{ia}-1,\qquad \lambda_i=\lambda_{i'}\Leftrightarrow i=i';\]
\item \[\Phi_{ial}(x)=\sum_{s=0}^l\sum_{t=1}^{g_{i,s+\nu_{ia}-l-1}}\!\!\!\!\!\!\varphi_{ia,l-s,t}(x)X_{its},\qquad\Psi_{ial}(x)=\sum_{s=0}^l\sum_{t=1}^{g_{i,s+\nu_{ia}-l-1}}\!\!\!\!\!\!\psi_{ia,l-s,t}(x)X_{its},\]

\noindent where

\renewcommand{\labelenumii}{\rm{(\theenumii)}}
\begin{enumerate}
\item $X_{ial}$, $i=1$, \dots, $m$, $a=1$, \dots, $g_i$, $l=0$, \dots, $\nu_{ia}-1$ is a complete set of constant eigen- and associated vectors of the matrix $A$, so that
\[AX_{ia0}=\lambda_iX_{ia0},\qquad(A-\lambda_iI_n)X_{ial}=X_{ia,l-1},\]
\[ i=1,\ldots, m,\quad a=1,\ldots,g_i,\quad l=1,\ldots, \nu_{ia}-1;\]
\item $g_{il}$ is the total number of vectors $X_{ial}$ with fixed $i=1$, \dots, $m$ and $l=0$, \dots, $\nu_{i1}$, so that
\[\sum\limits_{l=0}^{\nu_{i1}}g_{il}=\sum\limits_{a=1}^{g_i}\nu_{ia}\]
is the algebraic multiplicity of the eigenvalue $\lambda_i$ of the matrix $A$, $i=1$, \dots,~$m;$
\item $\varphi_{ialb}(x)$ and $\psi_{ialb}(x)$ for any $i$, $a$, $l$ and $b$ are a scalar functions$;$
\end{enumerate}
\item the Wronskian of all vector-functions $\Phi_{ial}(x)$ and $\Psi_{ial}(x)$ does not vanish on the entire axis.
\end{enumerate}}

In view of the latter criterion, relation \gl{vmp2.1} and the result of \cite{sokolov12,sokolov13} on constructing of a matrix intertwining operator from a set of associated vector-functions the following corollary is valid.\\

\noindent {\bf Corollary 1} (criterion of existence of a symmetry matrix for a matrix Hamiltonian).

\noindent{\it A constant $n\times n$ matrix $A$ is a symmetry matrix for an $n\times n$ matrix Hamiltonian~$H$ of Schr\"odinger form, $[H,A]=0$,

\vskip0.5pc

\noindent if and only if

\vskip0.5pc

\noindent there is a double set of associated vector-functions
\[\Phi_{ial}(x),\quad\Psi_{ial}(x),\qquad i=1,\ldots, m,\quad a=1,\ldots,g_i,\quad l=0,\ldots, \nu_{ia}-1,\] \[\sum_{i=1}^m\sum_{a=1}^{g_i}\nu_{ia}=n,\qquad \nu_{i1}\geqslant\nu_{i2}\geqslant\ldots\geqslant\nu_{i,g_i},\forall i\]
of the Hamiltonian $H$ such that this set satisfies conditions $1$ -- $3$ of the Criterion of partial strong minimizability from the right of a matrix intertwining operator.}

\vskip0.5pc

The following Section \ref{sec6} contains more useful in practice sufficient condition of existence of a constant symmetry matrix for a matrix Hamiltonian.

\section{Existence of a constant symmetry matrix for a matrix Hamiltonian\la{sec6}}

 Let us formulate the sufficient condition of existence of a constant symmetry matrix for a matrix Hamiltonian. This condition provide us with the opportunity to receive the  formula  for constructing of a matrix Hamiltonian with a given constant symmetry matrix which is useful in practice.

\vskip0.5pc

\noindent {\bf Sufficient condition of existence of a constant symmetry matrix for a matrix Hamiltonian.}

 \noindent{\it A constant $n\times n$ matrix $A$ is a symmetry matrix for an $n\times n$ matrix Hamiltonian $H$ of Schr\"odinger form, $[H,A]=0$, if there is a set of associated vector-functions
\[\Phi_{ial}(x),\qquad i=1,\ldots, m,\,\,\, a=1,\ldots,g_i,\,\,\, l=0,\ldots, \nu_{ia}-1,\quad \sum_{i=1}^m\sum_{a=1}^{g_i}\nu_{ia}=n,\]
of the Hamiltonian $H$ such that
\renewcommand{\labelenumi}{\rm{(\theenumi)}}
\begin{enumerate}
\item this set satisfies conditions $1$ and $2$ of the Criterion of partial strong minimizability from the right of a matrix intertwining operator$;$
\item the Wronskian of all vector-functions $\Phi_{ial}(x)$ does not vanish on the entire axis.
\end{enumerate}
}

\vskip0.5pc

This condition allows us to use the following method of constructing of a matrix Hamiltonian with a given constant symmetry matrix.

Arbitrary $n\times n$ matrix Hamiltonian $H$ of Schr\"odinger form with a given constant symmetry $n\times n$ matrix~$A$ can be found with the help of the following formula,
\[H=-\partial^2I_n+A+{\bf\Phi}''(x){\bf\Phi}^{-1}(x),\]
where ${\bf\Phi}(x)$ is an $n\times n$ matrix-valued function constructed from $n$ vector-functions
\[\Phi_{ial}(x),\quad i=1,\ldots, m,\,\,\, a=1,\ldots,g_i,\,\,\, l=0,\ldots, \nu_{ia}-1,\quad \sum_{i=1}^m\sum_{a=1}^{g_i}\nu_{ia}=n,\]
 as from columns. In this case a scalar functions $\varphi_{ialb}(x)$ from decompositions
\[\Phi_{ial}(x)=\sum_{s=0}^l\sum_{t=1}^{g_{i,s+\nu_{ia}-l-1}}\!\!\!\!\!\!\varphi_{ia,l-s,t}(x)X_{its}\]
(see above) are arbitrary parameterizing functions such that the Wronskian of all vector-functions $\Phi_{ial}(x)$, {\it i.e.} $\det{\bf\Phi}(x)$, does not vanish on the entire axis.

Let us elucidate the advantages of the above construction with two examples of constructing of arbitrary $2\times2$ matrix Hamiltonian with a given symmetry matrix $A$. In the first of these examples there are two different eigenvalues for the matrix $A$ and in the second of these examples normal (Jordan) form of $A$ is a Jordan block.

\noindent {\bf Example 1:} Constructing of a matrix $2\times 2$ Hamiltonian with a given symmetry matrix $A$ that has two different eigenvalues.

If $A$ is a given $2\times 2$ matrix that has two different eigenvalues,
\[AX_i=\lambda_iX_i,\qquad\lambda_i\in{\Bbb C},\quad X_i=\bigg(\begin{matrix}x_{i1}\\x_{i2}\end{matrix}\bigg) \ne\bigg(\begin{matrix}0\\0\end{matrix}\bigg),\quad i=1,2,\qquad\lambda_2\ne\lambda_1,\]
then arbitrary $2\times2$ matrix Hamiltonian $H$, for which $A$ is a symmetry matrix, can be found as follows:
\[\Phi_i(x)=\varphi_i(x)X_i,\quad i=1,2,\qquad {\bf\Phi}(x)=\bigg(\begin{matrix}\varphi_1(x)x_{11}&\varphi_2(x)x_{21}\\ \varphi_1(x)x_{12}&\varphi_2(x)x_{22}\end{matrix}\bigg),\]
\[W[\Phi_1(x),\Phi_2(x)]\equiv\det{\bf\Phi}(x)=\varphi_1(x)\varphi_2(x)\Delta,\qquad \Delta=x_{11}x_{22}-x_{12}x_{21},\]
\begin{equation}H=-\partial^2I_2+{1\over\Delta}\Big({{\varphi''_1}\over{\varphi_1}}+\lambda_1\Big)
\begin{pmatrix}x_{11}x_{22}&-x_{11}x_{21}\\x_{12}x_{22}&-\!x_{12}x_{21}\end{pmatrix}+
{1\over\Delta}\Big({{\varphi''_2}\over{\varphi_2}}+\lambda_2\Big)
\begin{pmatrix}-x_{12}x_{21}&x_{11}x_{21}\\-x_{12}x_{22}&x_{11}x_{22}\end{pmatrix},\label{H1}\end{equation}
\begin{equation}A={\lambda_1\over\Delta}\bigg(\begin{matrix}x_{11}x_{22}&-x_{11}x_{21}\\x_{12}x_{22}&-x_{12}x_{21}\end{matrix}\bigg)
+{\lambda_2\over\Delta}\label{A1}\bigg(\begin{matrix}-x_{12}x_{21}&x_{11}x_{21}\\-x_{12}x_{22}&x_{11}x_{22}\end{matrix}\bigg),\end{equation}
where $\varphi_1(x)$ and $\varphi_2(x)$ are arbitrary smooth scalar functions without zeroes. The fact that $[H,A]=0$ follows from evident commutativity of all matrices from the right-hand parts of (\ref{H1}) and (\ref{A1}).

\vskip0.5pc

\noindent {\bf Example 2:} Constructing of a matrix $2\times 2$ Hamiltonian with a given symmetry matrix $A$ that has eigen- and associated vectors.

If $A$ is a given $2\times 2$ matrix that has eigen- and associated vectors,
\[AX_0=\lambda_0X_0,\quad(A-\lambda_0I_2)X_1=X_0,\qquad\lambda_0\in{\Bbb C},\quad X_i=\bigg(\begin{matrix}x_{i1}\\x_{i2}\end{matrix}\bigg)\ne\bigg(\begin{matrix}0\\0\end{matrix}\bigg),\quad i=0,1,\]
then arbitrary $2\times2$ matrix Hamiltonian $H$, for which $A$ is a symmetry matrix, can be found as follows:
\[\Phi_0(x)=\varphi_0(x)X_0,\quad\Phi_1(x)=\varphi_0(x)X_1+\varphi_1(x)X_0,\quad {\bf\Phi}(x)=\bigg(\begin{matrix}\varphi_0x_{01}&\varphi_0x_{11}+\varphi_1x_{01}\\\varphi_0x_{02}&\varphi_0x_{12}+
\varphi_1x_{02}\end{matrix}\bigg),\]
\[W[\Phi_0(x),\Phi_1(x)]\equiv\det{\bf\Phi}(x)=\varphi^2_0(x)\Delta,\qquad \Delta=x_{01}x_{12}-x_{02}x_{11},\]
\begin{equation}H=-\partial^2I_2+\Big({{\varphi''_0}\over{\varphi_0}}+\lambda_0\Big)I_2+{1\over{\Delta}}
\Big({{\varphi''_0\varphi_1-\varphi_0\varphi''_1}\over{\varphi_0^2}}-1\Big)
\begin{pmatrix}x_{01}x_{02}&-x_{01}^2\\x^2_{02}&-x_{01}x_{02}\end{pmatrix},\label{H2}\end{equation}
\begin{equation}A=\lambda_0I_2-{1\over\Delta}\bigg(\begin{matrix}x_{01}x_{02}&-x_{01}^2\\x^2_{02}&-x_{01}x_{02}\end{matrix}\bigg).\label{A2}\end{equation}
where $\varphi_0(x)$ and $\varphi_1(x)$ are arbitrary smooth scalar functions and $\varphi_0$ is without zeroes. The fact that $A$ is a symmetry matrix for $H$ is obvious in view of (\ref{H2}) and (\ref{A2}).

\kern-0.5pc

\section{Conclusions\la{sec9}}
Let us outline some possible problems for future studies.
\renewcommand{\labelenumi}{\rm{(\theenumi)}}
\begin{enumerate}
\item To find a criterion of complete strong minimizability for a matrix intertwining operator.
\item To elaborate methods of spectral design for matrix Hamiltonians with the help of matrix intertwining operators of arbitrary order and, in particular, to reveal a criterion for transformation vector-functions that provides required changes for the spectrum of the corresponding final matrix Hamiltonian with respect to the spectrum of an initial matrix Hamiltonian. For this purpose one could try  to generalize Index Theorem and Lemma 4 of \cite{ancaso07,sokolov07np} for the matrix case.
\item To study (in)dependence of matrix differential intertwining operators in the way analogous to one of \cite{anso03} and, in particular, to define the notions of dependence and independence for these operators, to find a criterion of dependence for them and to solve the questions on maximal number of independent matrix differential intertwining operators and on a basis of such operators.
\item By analogy with \cite{anso03,anso09} to investigate  properties of a minimal matrix differential hidden symmetry operator.
\item To study (ir)reducibility of matrix differential intertwining operators and, in particular, to classify irreducible and absolutely irreducible \cite{sokolov13} matrix differential intertwining operators in the way analogous to one of \cite{acdi95,anca04,samsonov99,anso06,sokolov07,sokolov10,ferneni00,tr89,dun98,khsu99,fermura03,fermiros02',fersahe03,samsonov06}.
\end{enumerate}

\section*{Acknowledgments}
This work was supported by RFBR grant 13-01-00136-a. The authors acknowledge Saint-Petersburg State University for a research grant 11.38.660.2013. AVS is grateful also to organizers of PHHQP~12 for hospitality and to Saint-Petersburg State University for a travel grant 11.46.1849.2013.

\end{document}